\documentclass[aip,graphicx,reprint]{revtex4-1}
\pdfoutput=1
\usepackage{color}
\usepackage{amssymb,amsmath}
\usepackage{subfigure}
\usepackage{epstopdf}
\usepackage{graphicx}

\usepackage[tikz]{bclogo}
\usetikzlibrary{shapes,snakes}
\makeatletter
\def\ScaleIfNeeded{%
\ifdim\Gin@nat@width>\linewidth
\linewidth
\else
\Gin@nat@width
\fi
}
\makeatother

\definecolor{mma1}{RGB}{63,60,153}
\definecolor{mma2}{RGB}{135,60,121}
\definecolor{mma4}{RGB}{61,134,99}
\definecolor{mma3}{RGB}{135,123,79}

\begin{document}

\title{Quantum Origins of Molecular Recognition and Olfaction in Drosophila} 

%

\author{Eric R. Bittner}\affiliation{Depts.  of Chemistry and Physics, University of Houston, Houston TX, 77204}
\author{Adrian Madalan}\affiliation{Dept.  of Biology and Biochemistry, University of Houston, Houston TX, 77204}
\author{Arkadiusz Czader}\affiliation{Dept.  of Chemistry,  University of Houston, Houston TX, 77204}
\author{Gregg Roman}\affiliation{Dept.  of Biology and Biochemistry, University of Houston, Houston TX, 77204}

\begin{abstract}
The standard model for molecular recognition of an odorant is that receptor sites discriminate by 
molecular geometry as evidenced that  two chiral molecules may smell very differently. 
However, recent studies of isotopically labeled olfactants indicate that there may be a
molecular vibration-sensing component to olfactory reception, specifically in the spectral region around
 2300 cm$^{-1}$.  Here we present a 
donor-bridge-acceptor model for olfaction which attempts to explain this effect.  Our model, based
upon accurate quantum chemical calculations of the olfactant (bridge) in its neutral and ionized states,
posits that internal modes of the olfactant are excited impulsively 
during hole transfer from a donor to acceptor site on the receptor, 
specifically those modes that are resonant with the tunneling gap. 
By projecting the impulsive force onto the internal modes, 
we can determine which modes are excited at a given value of the donor-acceptor tunneling gap. 
Only those modes resonant with the tunneling gap and are impulsively excited will give a significant contribution to the 
inelastic transfer rate.  Using acetophenone as a test case, our
 model and experiments on {\em D. melanogaster} suggest that isotopomers of a given olfactant 
give rise to different odorant qualities.  These results support the notion 
that inelastic scattering effects may play a role in discriminating 
between isotopomers
but that this is not a general spectroscopic effect. 
\end{abstract}
\maketitle


\section{Introduction}
The general model for  detection is that the  response is triggered by the transfer of an 
electron from a  donor (D) to an  acceptor (A) within the receptor site 
by the presence of an olfactant molecule that provides a bridge between the two. In the absence of the 
odorant, the distance between D and A is too great and electron transfer is inefficient.
Placing an olfactant (B) between the two allows the electron transfer to occur either as a single coherent 
scattering event from D to A, or as a sequence of two incoherent hops, first from D to B then from B to A. 
This simple model is consistent with the standard ``swipe-card'' model for odor detection since implicit in this 
is that B must fit into some sort of pocket and be in a correct alignment between D and A in order for the 
charge transfer process to occur.  



\begin{figure*}[t]
\subfigure[]{\includegraphics[width=0.5\ScaleIfNeeded]{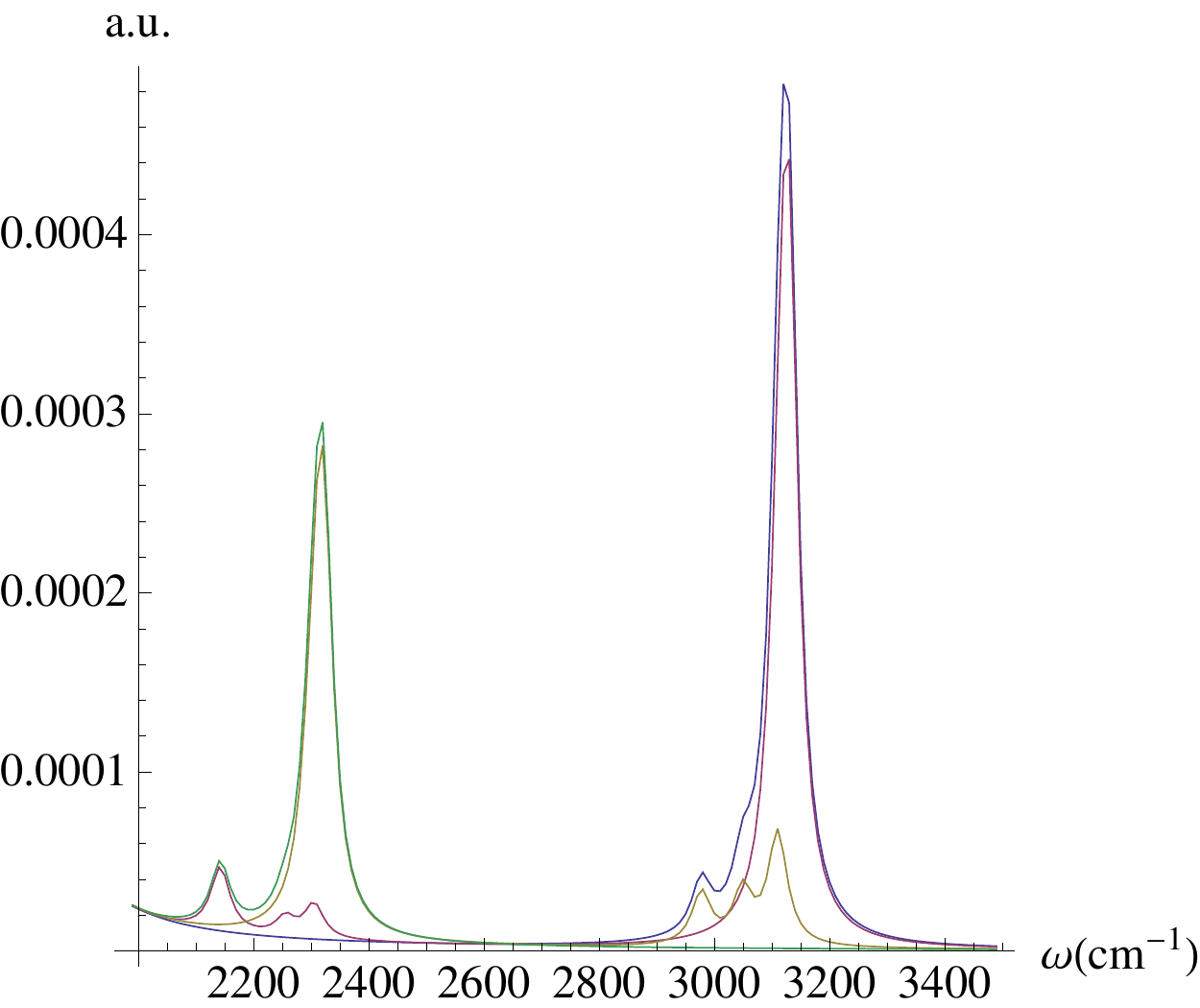}\label{acp-spect}}
\subfigure[]{\includegraphics[width=0.5\ScaleIfNeeded]{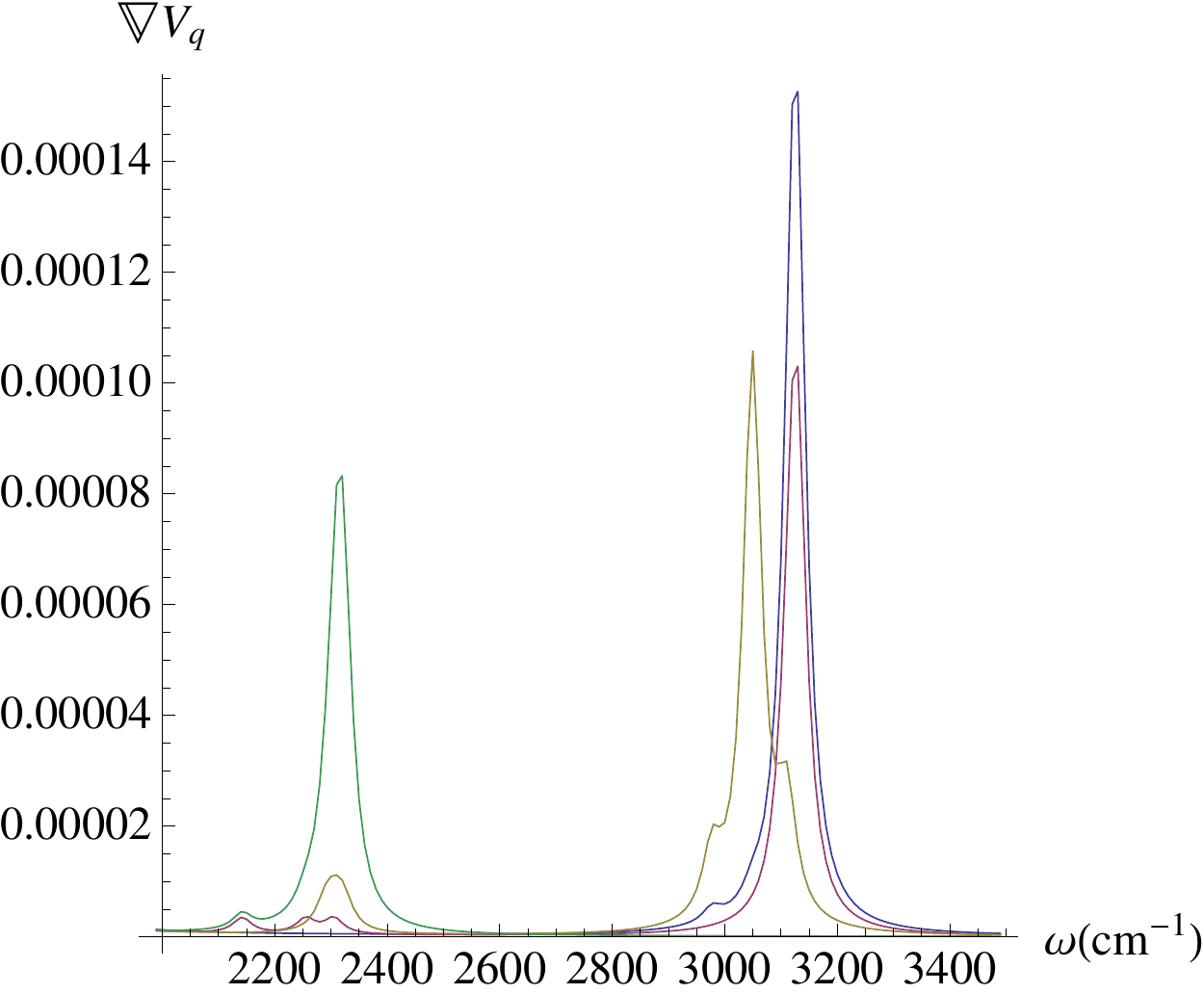}\label{acp-grad}}
\caption{Calculated IR Spectrum  of neutral (a) and  projection of the energy gradients (in atomic units) (b) of singly oxidized ACP and its deuterated isotopomers in the 2000-3500 cm$^{-1}$ C-D  and C-H stretching regions.
Color key: 
${\color{mma1}-}$: hACP, 
${\color{mma2}-}$: d3ACP,
${\color{mma3}-}$: d5ACP, 
${\color{mma4}-}$: d8ACP.
}
\end{figure*}
An interesting extension to this paradigm is that molecular shape may not be the sole 
deciding factor in scent recognition.
Indeed, it has been observed recently that fruit flies ({\em D. melanogaster}) can na\"ively discriminate between several isotopomers that have deuteriums substituted for hydrogen\cite{Franco01032011}.    These flies can also be trained to associate a specific odorant with an electric shock resulting in the specific avoidance of the conditioned odorant\cite{Tully:1985kx}. Flies trained in this manner can discriminate deuterated from hydrogenated isotopomers, indicating an ability to perceive differences in these odorants.   
The ability to discriminate between the h-ACP and the d8-ACP has been shown to require the Orco (Or83b) olfactory co-receptor \cite{Franco01032011,Larsson2004703}.   
Mutant flies that lack the  gene are broadly anosmic and fail to discriminate between isotopomers, indicating the behavioral discrimination of isotopomers relies on olfactory perception. 
In addition, there is a curious observation that fruit flies trained to discriminate deuterated olefins also discriminate non-deuterated olfactants with strong IR peaks in the 2300 cm$^{-1}$ range.\cite{Franco01032011,Joachim:541,Turin01121996,TURIN2002367}
Lastly, using the negatively-reinforced learning paradigm, Drosophila has been shown capable of generalizing a trained odorant to another that shares a similar vibrational spectrum.  For example, when deuterated d5-benzaldehyde is paired with electric shocks, the flies will subsequently avoid fully-deuterated d17-octanol 
(vs. undeuterated h-octanol),
suggesting that the C-D vibrational mode is a salient feature of odorant perception in this species.   


A controversial and speculative explanation of these observations 
 is that there is an spectroscopic component to 
olfaction.
\cite{Turin01121996,TURIN2002367}
This model supposes that the bridge facilitates inelastic electron scattering and as such the olfactory response 
can be predicted by comparing infra-red (IR) spectra of various olfactants. 
Isotopic substitution of H for D shifts the CH stretching frequency from  the  2850 -  3100 cm$^{-1}$ range into the 2300 cm$^{-1}$ range. 
Few molecules absorb in this IR region  and there is little or no biological need or evolutionary pressure that 
we know for detecting deuterated compounds.  Needless to say, 
this theory has been met with considerable skepticism since it runs contrary to more commonly held model in  which the geometric shape and chemical nature of the olfactant are the primary components of olfactory reception. \cite{Hettinger02082011}
Moreover, many of the claims in Ref. \onlinecite{Turin01121996} and Ref. \onlinecite{TURIN2002367} 
were show to be inconsistent with psychophysical tests 
performed on 
human subjects by Keller and Vosshall. \cite{Keller:2004fk} 
For example,  in this study human subjects could not discriminate by smell between 
deuterated and non-deuterated forms of acetophenone.  On the other hand, the molecular mechanism for 
olfaction in humans might be different since humans have far fewer active odor receptor genes than 
other mammals and insects. 

%

\section{Theoretical Model}

While vertebrate olfactory receptors are G-protein coupled, insect ORs appear to have a 
deferent structure and act as ligand-gated cation channels. 
The Or83b odorant receptor gene is a broadly expressed receptor  in drosophila and is remarkably conserved 
amongst the insect species.  
Information concerning the molecular structure at the olfactant binding site remains illusive, 
may involve odorants binding to  Cu(II) and Zn(II) ions bound to the protein loop, activating the 
G protein through a structural  
a ``shuttle-cock'' mechanism that leads to cascade of events that eventually leads 
to some neural activity. \cite{Wang:2003}

%
Since very little is 
known about the molecular level details of the binding of the olfactant to a receptor site, our
goal here is to piece apart the possible contributions to the transfer rate,  determine 
whether or not a given bridge molecule can have an inelastic component to the rate, and give at best 
an indication of the relative strengths of the inelastic components. 
However, for a given 
chemical olfactant, it is unlikely that discrimination between isotopomers can be a purely electronic effect
 since isotopomers have  identical electronic structures within the Born-Oppenheimer approximation. 
It is unlikely that the discrimination could be differences in mass since this would account for at
most a few percent in terms of the diffusivity. 
For example, in comparing the diffusion constant of 
acetophenone to its fully deuterated isotopomer, one has at best a 3\% difference. 
Also, at ordinary physiological temperatures, there is no significant vibrational contribution to the heat capacity coming 
from the relevant C-H or C-D stretching or bending modes at 3100 cm$^{-1}$ and 2300 cm$^{-1}$, respectively. 

We  assume that  at the heart of the process, a charge is transferred
from a donor  to an acceptor  within the receptor site itself.  
In the absence of an odorant, the distance between the 
donor and acceptor is too great to allow efficient electron transfer.  When the odorant is present, it acts like a bridge between the donor and acceptor allowing for more efficient electron tunneling between the two. \cite{Devault:1984fk} 
Aromatic compounds are typically better electron donors ({\em i.e.} hole receptors) due to their electron-rich $\pi$ -system.  With this in mind, we 
shall assume that the bridge/olfactant acts as an intermediate for hole transfer between a donor site and an acceptor site. 
Energy conservation requires that the energy transferred to vibrational motion is equal to the tunneling gap.  

For a donor-bridge-acceptor (DBA) system in the non-adiabatic limit of electron transfer theory, 
 the golden rule rate is given by 
 $$
 k_{da} = \frac{2\pi}{\hbar} | V_{da}|^2 {\cal F}(E_{da}).
 $$
There are two contributing factors to the transfer rate.   First, 
$V_{da}$ is the effective coupling between donor and acceptor which depends upon the 
electronic coupling between the donor and acceptor and ${\cal F}(E_{da})$ is the thermally averaged Franck-Condon weighted 
 density of nuclear vibrational states between the donor and acceptor and $E_{da}$ is the energy gap between donor and acceptor. 

\begin{widetext}
\subsection{Inelastic scattering model}
Before describing our approach, let us briefly review the model by Lambe and Jakelvic for molecular vibrational spectroscopy via
inelastic electron tunneling in a tunnel junction system. \cite{PhysRev.165.821}
 In a tunneling device, where there is a barrier potential 
$U(z)$ separating two metallic leads, the WKB approximation give the electronic part of the tunneling matrix element as 
\begin{eqnarray}
|V_{da}| \propto \exp\left[-\int_{0}^{L}\left(\frac{2m}{\hbar^{2}}\right)(U(z) + U_{int}(z) - (E-E_{\perp}))^{1/2} dz\right]
\end{eqnarray}
where $L$ is the spacing between leads (i.e. the spatial thickness of the barrier), $E$ is the total electronic energy, $E_{\perp}$ is the kinetic energy associated with motion perpendicular to the barrier, and $U_{int}(z)$ is small perturbing potential due to the presence of an impurity 
molecule in the tunneling region.  
\end{widetext}
In a classic paper by Scalapino and Marcus\cite{Scalapino:1976fk}, they considered the the case where 
a molecule with a permanent dipole moment $\mu_{o}$ is located close to  one of the electrodes so that its image dipole must be included.  This 
leads to an interaction potential between the passing electron and dipole of the form
\begin{eqnarray}
U_{int}(z) = 2e \mu_{o}z/(z^{2} + r_{\perp}^{2})^{3/2}
\end{eqnarray}
where $r_{\perp}$ is the distance from the molecule and the electrode and $\mu_{o}$ is the 
molecular dipole operator.  For the case of a single molecule in the tunneling region
the current couples to the the dipole oscillations of the molecule leading to an inelastic contribution 
that contains the vibrational transition moment.   Lambe and Jakelvic also show that even in when the molecules in the tunneling gap 
lack permanent dipole moments, the current can couple to the polarizibility  of the molecules leading to Raman contributions to the inelastic current.  Here, again one considers the interaction between the impurity molecule, the passing electron, and the nearest image dipole to 
show that 
\begin{eqnarray}
U_{int}(z) = -4e^{2}\alpha z^{2}/(z^{2} + r_{\perp}^{2})^{3}
\end{eqnarray}
where $\alpha$ is the polarizability of the molecule.  In this case, the scattering of the electron induces Raman-like transitions within the
impurity molecule.  Both cases lead to golden rule expressions for the inelastic contributions to the current-voltage curves 
in the presence of molecular impurities in the system.
Brookes {\em et al.}\cite{PhysRevLett.98.038101}  follow a similar lines and arrive at estimates for whether or not an inelastic
tunneling component would be possible to observe in a model parameterized by physiological considerations.

However we argue that such inelastic barrier tunneling models are not suitable for the case at hand where
 we have electron transfer between localized states on the donor, bridge, and acceptor.
While connection between 
 the electron transfer  rate and the zero-bias molecular conduction was discussed by Nitzan \cite{jp003884h}
 in comparing the Landauer formula to the Marcus rate, the relation is established when 
 the orbitals of the bridging molecule are in contact with orbitals of the donor and acceptor leads. 
This is not the case in the model described above.

 \subsection{Donor-Bridge-Acceptor model}
 
Since the electron transfer  rates are likely far slower than the structural responses they trigger, 
the central goal in this paper is to determine those internal vibrational modes of the bridge species that 
are important in accommodating the inelastic scattering rather than 
providing actual transition rates.
 Bridge mediated charge transfer is a broadly studied topic\cite{marcus:599,marcus:679,marcus:966,Marcus:1985fk,medvedev:3821,PhysRevE.81.027101,tanaka:11117}, 
and we start by assuming that  we have three relevant sets of diabatic states denoted by 
$|\psi_{d} \rangle = |D^{+}BA\rangle$, 
$|\psi_{b} \rangle = |DB^{+}A\rangle$, 
and
$|\psi_{a} \rangle = |DBA^{+}\rangle$  corresponding to the initial, intermediate, and final quantum states of the system. 
In $|\psi_{d} \rangle$, the charge is localized in a donor orbital and the bridge is in its neutral electronic state.
In $|\psi_{b} \rangle$, the bridge is in a singly oxidized state, lastly in $|\psi_{a} \rangle$, the bridge is again in a 
neutral state and the charge is localized in an orbital on the acceptor.
Within a diabatic picture, we have the following  Hamiltonian: 
\begin{eqnarray}
H =\left[
\begin{array}{ccc}
\hat H_{d}  &  \hat J_{db}  &0 \\
\hat J_{bd}   & \hat H_{b}  & \hat J_{ba} \\
0   & \hat J_{ab} & \hat H_{a}
\end{array}
\right]
\end{eqnarray}
where each term on the diagonal represents the electronic + nuclear Hamiltonians for each state. 
$\hat J_{bd}$ and $\hat J_{ba}$ are the electronic interactions between bridge and donor or acceptor states respectively. 
We assume that the the through-space coupling between the donor and acceptor are  negligible  
compared to the other couplings in the system. \cite{C2CP40579B}

Let us re-cast this as a reduced two state problem using the Feshbach method so that 
the transition from the donor to acceptor states is via resonant scattering involving the 
bridge (olfactant) molecule.\cite{Joachain:1975xi,child,cr00005a007,Bixon:1999kx}
The results in an effective Schr\"odinger equation 
\begin{eqnarray}
H_{eff} = \hat H_{d} +\hat H_{a}+ {\cal V}_{eff}(E)\end{eqnarray}
where $ \hat H_{d} +\hat H_{a}$ is the unperturbed  system lacking the bridge and 
\begin{eqnarray}
{\cal V}_{eff}(E)  =  \hat J_{ab}  \frac{1}{(E-\hat H_{b})} \hat J_{bd}  \label{veff}
\end{eqnarray}
is an {\em effective } coupling matrix element between the donor and acceptor due to the presence of the 
bridging molecule which depends upon the scattering energy, $E$. \cite{C2CP40579B,cr00005a007,Bixon:1999kx}
This operator has a series of poles located at the eigen-energies of $H_{B}$ and 
contains both resonant and non-resonant components. 


At the heart of Eq.~\ref{veff}  is the Green's function for the evolution 
of a nuclei on  the potential energy
surface of the ionized species. We can imagine this in the 
dynamical sense: Upon ionization, the nuclei in $B$ experience a sudden change in their electronic environment corresponding to the 
charge transfer from the donor to the bridge. This creates a vibrational wave function on the 
potential energy surface of the ionized species centered about the ground-state nuclear geometry.  We can write this 
as
\begin{eqnarray}
{\cal V}_{eff}(E) &=& \langle \psi_{a} | \hat J_{ab}G_{b}(E)\hat J_{bd}|\psi_{d}\rangle \nonumber \\
&=& \frac{J_{ab}J_{bd}}{i\hbar}\int_{0}^{\infty} e^{-iEt/\hbar}C_{da}(t) dt \label{cot}
\end{eqnarray}
where $C_{da}(t)$ is correlation function for the propagation a vibrational wave packet 
 on the Born-Oppenheimer potential, $V^{(+)}$, corresponding to this new electronic configuration, then projected onto the 
 manifold of vibrational states of the final electronic configuration, where as above we partition the electronic contribution from the 
 nuclear dynamics, except that our time evolution occurs on the $V^{(+)}$ potential corresponding to the bridging state. 
 This gives  the correlation function as
 $$
 C_{da}(t) = \sum_{a}\langle n_{a} | e^{-iH_{b}t/\hbar} | 0_{d}\rangle  = \sum_{a} \langle n_{a} | \psi(t) \rangle
 $$
 where we have assumed the initial wave packet to the be the ground-state vibrational wave function and the sum is 
 over final vibrational states $|n_{a}\rangle$.
 Upon promotion to the ionized state, the nuclei receive an impulsive force along the direction of the gradient of $V^{(+)}$.  
This is certainly the case for a  classical oscillator displaced
from its equilibrium position $q_o$ to $q'$ where the force acting on the particle is $-k(q'-q_{o})$.  
The oscillator then follows an elliptical trajectory in phase-space.
Secondly, since the wave function is simply the displaced harmonic oscillator ground-state wave function and the well is harmonic, the shifted state evolves as a Glauber coherent state in  phase space without spreading or contracting.  
Thus, we approximate the time correlation function as
\begin{eqnarray}
C_{da}(t) &=& \sum_{j}  e^{i\omega_{j} t} \langle n_{j}|e^{ip_{j}x} |0_{d}\rangle \nonumber \\
&\approx& \sum_{j}  e^{i\omega_j t} i p_{j} \mu_{j}/e 
\end{eqnarray}
where $p_{j}$ is the momentum imparted along normal coordinate $j$,  $\omega_{j}$ is the normal mode frequency, 
and $\mu_{j} = e\langle 1_{j}| x |0_{d}\rangle $  is the transition dipole moment for the $j$th vibrational transition.

We can now evaluate Eq.~\ref{cot} by taking the Fourier transform of the correlation function. 
This gives the effective potential as a series of $\delta$-functions 
 \begin{eqnarray}
{\cal V}_{eff}(E) \approx
{J_{ab}J_{bd}}
\sum_a  p_{b}(\omega_{a}) \langle 1_{a}| x |0_{d}\rangle  \delta(E_{da} - \hbar\omega_{a})
\label{cot2}
\end{eqnarray}
and we recall that the 
 delta-function carries units of inverse-energy. 
Energy conservation requires that total momentum imparted 
 to all the oscillators be such that $E_{da} = \sum_{n}p_{n}^{2}/2$ in mass-scaled units.
From above, we assumed that the  momentum transferred to the mode  is proportional to the energy gradient of the ionized species 
along that direction evaluated at the equilibrium position of the neutral. 
This implies that to a first approximation the inelastic scattering of an electron via the bridge species excites the vibrational modes of 
 the bridge that are also infra-red (IR) active.  This is a central component of Turin's spectroscopic theory of olfaction.\cite{Turin01121996,TURIN2002367,Franco01032011}
However, here we see that {\em only those modes that are directed along  
$\vec\nabla V^{(+)}$ will be excited by the impulsive scattering process.  }

%

%



 Eq.~(\ref{cot2}) gives us a direct way to rapidly screen whether or not a given
 odorant is expected to exhibit an isotope effect using quantum chemical means.  
Starting from the equilibrium position of the neutral, one first determines the vibrational 
normal modes and frequencies of a given olfactant.  This gives the IR response of the molecule. 
We then determine the energy gradient of the ionized species at this geometry and project this onto the 
normal mode coordinates, 
which are  normalized linear combinations of the cartesian displacement coordinates for each atom in the molecule. 
Since $E_{da}$ remains an unknown in our model, 
we  require that this energy  be distributed amongst the normal coordinates in proportion to their projection onto the  energy gradient.
Thus, even if a mode has a strong IR response and satisfies energy conservation, 
unless that mode is  directly excited by the impulsive scattering process, then it will not contribute to the overall transition rate. 
Moreover, in order for CD stretches to play any role in the transfer rate, we have to
 make the assumption that $E_{da} \approx 2300 {\rm cm}^{-1}$.

\subsection{Numerical results}
As a test case, we consider acetophenone (ACP) and its deuterated isotopomers. In  d3-ACP, the hydrogens on the methyl group are replaced, in d5-ACP the hydrogens on the phenyl ring are replaced, and in d8-ACP all hydrogens are replaced with deuteriums.  In all cases, we  first perform geometry optimizations of the neutral species 
followed by calculations of vibrational frequencies using the B3LYP density functional with 6-31G(d) basis set in vacuum using the
NWChem quantum chemistry package.\cite{NWChem:2010} The same functional and basis set were then used to calculate the nuclear gradients and frequencies of the radical cation at the equilibrium geometry of the neutral. 
The gradients are then projected on to the normal mode eigenvectors giving the projection of the 
force onto a given normal mode with frequency giving us $p(\omega)$.  The normal mode calculations also give us the 
infra-red response, $\mu(\omega)$.  Since we know little about the actual binding site, we can only assume that the 
transfer integrals $J_{ab}$ and $J_{db}$ are independent of the vibrational modes and are the same for each isotopomer. 
  Figure \ref{acp-grad} shows the results of our calculations for the deuterated 
and non-deuterated forms of ACP along with the predicted infra-red responses. 
 This approach is numerically reliable and produces accurate potentials and gradients for this system without undue effort. Moreover, this general procedure can be applied to a broad set of olfactants.

Briefly, the peaks around 3100 cm$^{-1}$ correspond to the C-H stretching modes of the molecule.  
In the fully deuterated d8-ACP, this peak is shifted to 2300 cm$^{-1}$, 
consistent with the typical isotope shift of a C-D stretching mode.  
Furthermore, the IR spectra for all but the d8-ACP is void of peaks 
between 1800 - 3000 cm$^{-1}$.  Only the 
fully deuterated d8-ACP has any appreciable spectral density in this region. 
While there is little to note in comparing the intensities for the modes below 1800 cm$^{-1}$, 
the contribution to the gradient coming from the C-H and C-D stretching modes 
is remarkably strong. Most notable, the 
peak at 2300 cm$^{-1}$  is more or less the sum of contributions 
from the ring and methyl C-D stretching modes. 


Based upon the relative intensities of both the IR and energy gradients around 2300 cm$^{-1}$, 
the electron transfer model suggests that  
 {\em D.~melanogaster} should readily discriminate h-ACP and 
 d8-ACP. In comparing the predicted IR spectra, 
there is almost no IR oscillator strength in the C-D stretching modes 
for d3- and d5-ACP while d8-ACP exhibits a fairly strong signal.  However, in comparing the gradients,
the d5-ACP gradient has considerably smaller projection onto the 
C-D stretches modes and the d3-ACP projection is almost vanishingly small. 
Based upon gradients, one predicts that there should be an ability to differentiate between
h-ACP and d8-ACP. Moreover,  the flies  may be able to discriminate between
d3-ACP and d5-ACP even though both have very weak IR signals at 2300cm$^{-1}$. 

A crucial test of our model would be to identify an olfactant that can not be 
discriminated from its deuterated isotopomers.  For example, 
in removing an electron from ethylene, the C=C bond-length in ${\rm C}_{2}{\rm H}_{4}^{+}$ will be longer
 than in ${\rm C}_{2}{\rm H}_{4}$ (due to reduction in bond-order) but the C-H bond-lengths will be more or less unchanged.  
This suggests that the energy gradient in our model will be directed primarily 
along the C=C bond. 
Quantum chemical calculations of the sorts described above on ethylene and two of its isotopomers, 
${\rm HDCCH}_{2}$ and ${\rm D}_{2}{\rm CCH}_{2}$,
indicate the forces along the CD stretching coordinates these are far weaker in magnitude 
than along the CH stretching coordinates. More over they are about 40-fold weaker
than the corresponding forces in d8ACP but are only half of those in d3ACP.
However, the IR spectrum of ethylene does show an obvious isotope shift of the C-H stretching  
modes into the 2300cm$^{-1}$ region.
Consequently, our model would predict that 
while Drosophila can discriminate between hACP and d3ACP it is doubtful that they could  
discriminate between ethylene and its isotopomers within the limits of the 
error bars of the T-maze experiments. 

 \begin{figure*}
\subfigure[]{ \includegraphics[width=0.5\ScaleIfNeeded]{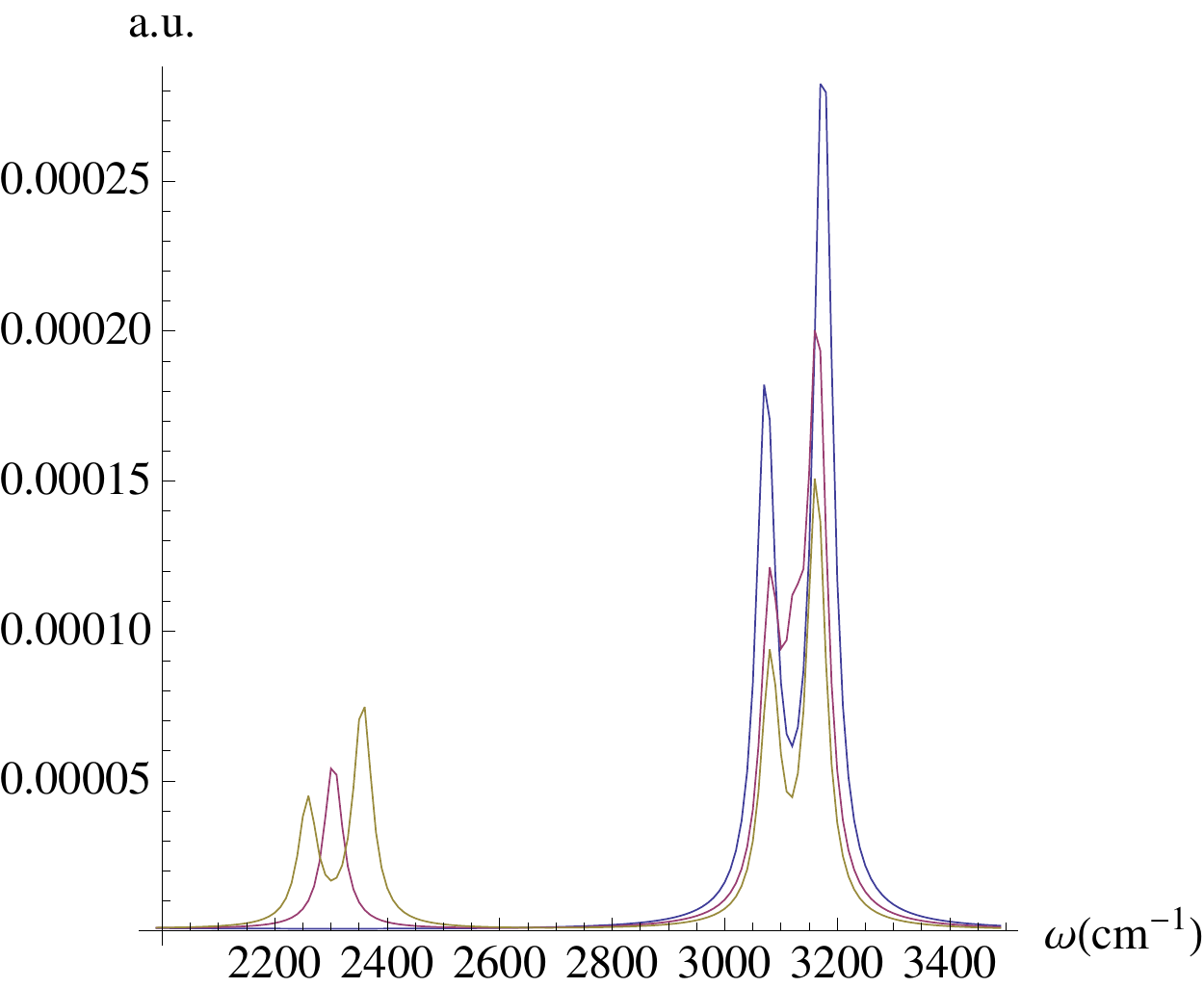}}
\subfigure[]{  \includegraphics[width=0.5\ScaleIfNeeded]{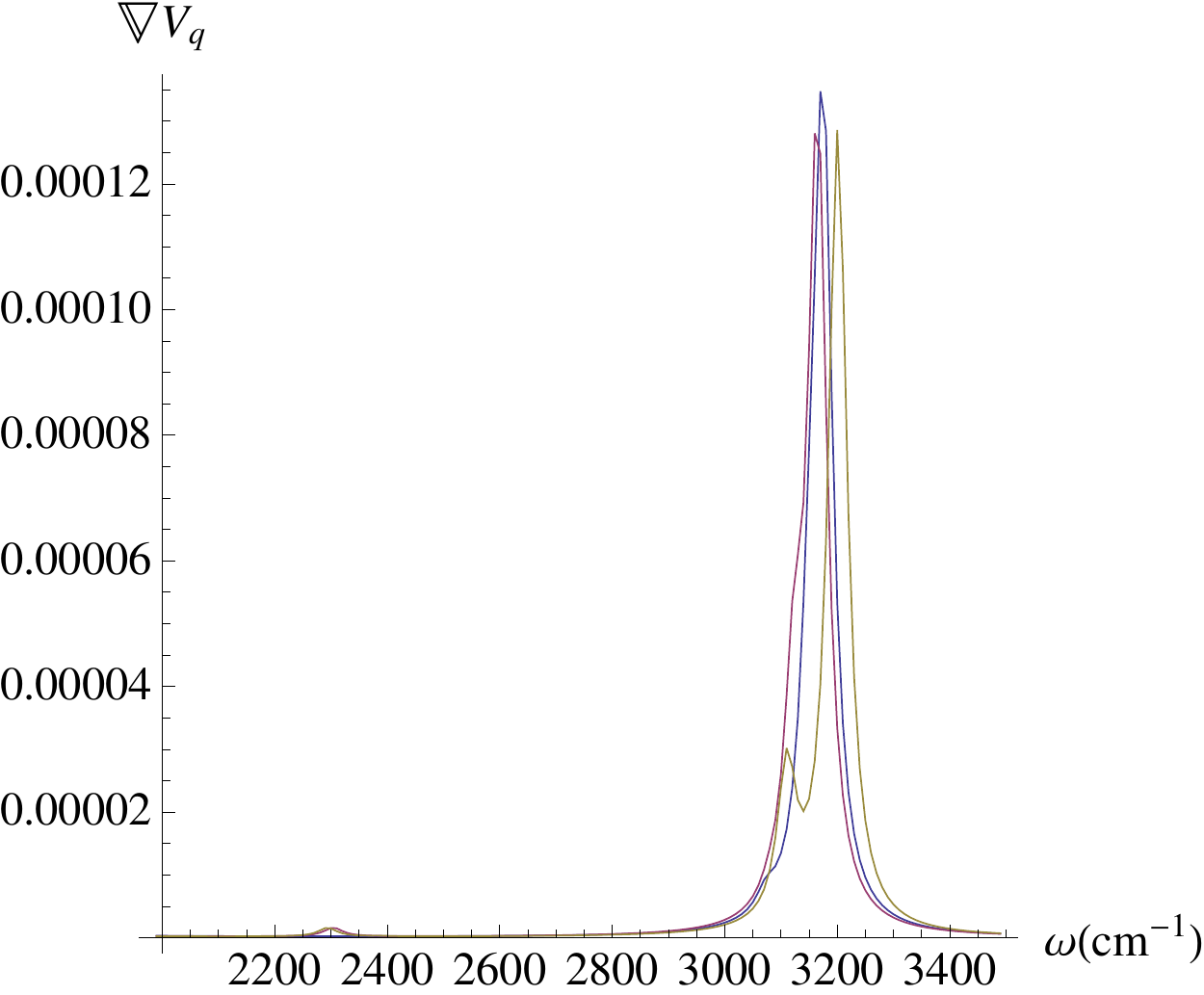}}
\caption{Computed IR Spectrum of ethylene (a) and projection of the energy gradients of singly oxidized ethylene (b) and its deuterated isotopomers in the 2000-3500 cm$^{-1}$ CD and CH stretching 
region. Color key: 
{\color{mma1}--} $C_{2}H_{4}$,
{\color{mma2}--} $DHC_{2}=C_{2}H_{2}$,
{\color{mma3}--} $D_{2}C_{2}=C_{2}H_{2}$}\label{ethylene}
 \end{figure*}

\section{Experimental Tests}

\begin{figure*}
\includegraphics[width=\ScaleIfNeeded]{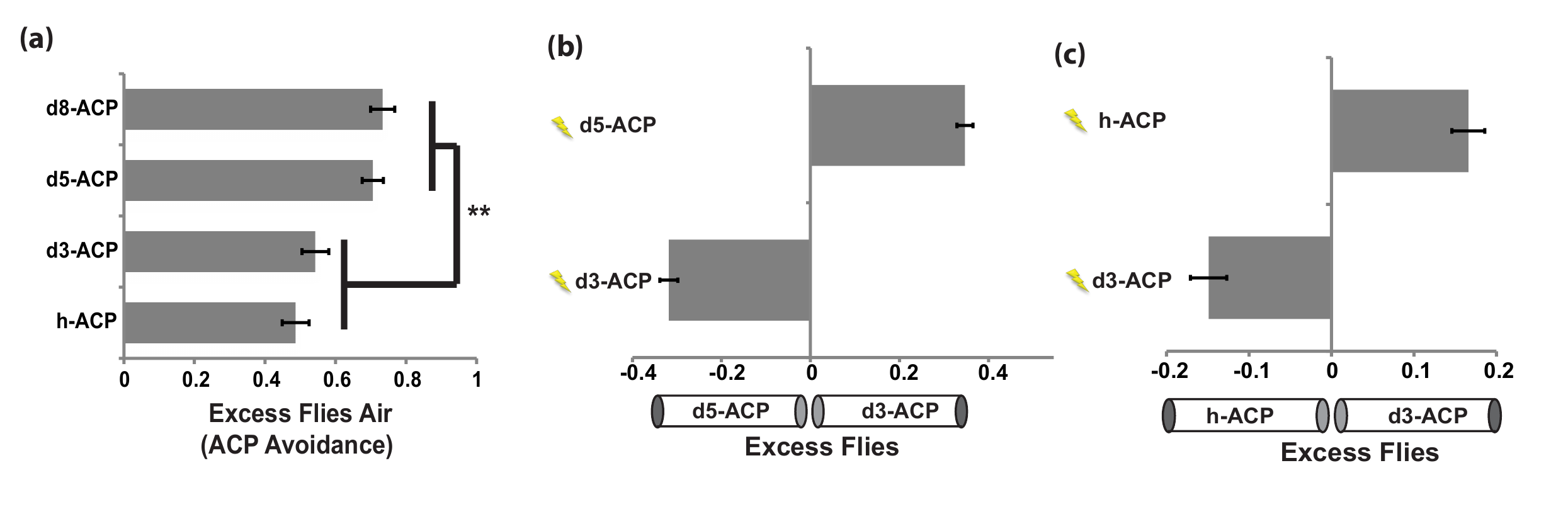}
\caption{  Olfactory Discrimination of Acetophenone isomers.   
a)  Drosophila melanogaster ($w^{1118}$) na\"ively avoid ACP in a T-maze.  The behavioral responses to both d5-ACP and d8-ACP are significantly stronger than to both the d3-ACP and h-ACP.  ($n = 22$ groups each)   
(b)  Drosophila can discriminate between d5-ACP and d3-ACP after training with electric shock.  During the training session, populations of flies are exposed to one of the  isotopomers  paired with an electric shock. When subsequently tested in the T-maze, the flies significantly prefer the unpaired odorant over the paired odorant ($p < 0.0001$; $n$ = 44 groups each).  
(c)Drosophila can discriminate between h-ACP and d3-ACP after training with electric shock. 
 }\label{GR2}
\end{figure*}

Previously, d5-ACP and d8-ACP odorants were found to be significantly more aversive to Drosophila than h-ACP \cite{Franco01032011}.  We verified this result, examining the na\"ive avoidance of Drosophila $w^{1118}$ to 0.3\% h, d3, d5, and d8-ACP using an olfactory T-Maze\cite{Tully:1985kx} (Figure~\ref{GR2}a).  The avoidance response to h-ACP and d3-ACP were indistinguishable (F3,80 = 11.724, Bonferonni-Dunn {\em post hoc} $p = 0.276$); the responses to d5 and d8 were also not significantly different from each other (p = 0.842).   However, the avoidance response to d5-ACP was significantly greater than the response to d3-ACP ($p = 0.0016$), suggesting deuteriums on the benzene ring provided a different odorant quality than the methyl  deuteriums, leading to an increased saliency for the d5-ACP and d8-ACP.   To test this hypothesis, we used negatively-reinforced olfactory learning \cite{Roman:2023fk}.   

Surprisingly, the $w^{1118}$ flies are capable of discriminating between d5-ACP and d3-ACP after training, indicating a significant difference in perceptual quality between these odorants (Figure~\ref{GR2}b; $t=24.714$, $p < 0.0001$).     
The $w^{1118}$  flies are also capable of discriminating between h-ACP and d3-ACP after training (Figure 3c; $t =10.596$; $ p< 0.001$).  The conditioned avoidance in this experiment was reduced in comparison to the conditioned avoidance of d5-ACP vs. d3-ACP, indicating  that d3-ACP is more similar in
 odorant quality to h-ACP than to d5-ACP.
The robust discrimination between these very closely related odorants is consistent with our model that predicts differences in the excitation of vibrational modes of the methyl and deuterium C-D bonds by the impulsive scatting.

\section{Summary}
It is clearly evident that {\em D.~melanogaster} can be trained to distinguish deuterated isotopomers of 
various olfactants.  Whether or not this is the result of a spectroscopic detection mechanism is an 
another question.  
Our model predicted that  Drosophila could be trained to discriminate between d8-ACP and d5-ACP, between d5-ACP and d3-ACP, and 
between d3ACP and hACP based upon comparison of the gradients at 2300cm$^{-1}$.  
This prediction was borne out in the training experiments. 
Moreover,  discrimination between d8-ACP and d5-ACP would
not have been predicted based upon {\em just} the IR spectrum. 
While deuteration does shift the C-H stretching mode into the otherwise empty
2300 cm$^{-1}$ region, our model suggests that 
{\em IR response alone does not determine whether or not the flies can discriminate isotopomers.}  
Moreover, this effect appears to be limited to a single spectral region and does not involve the 
entire IR spectrum.\cite{TURIN2002367,Turin01121996}

Lacking detailed molecular level knowledge of the 
receptor sites, we  speculate the tunneling gap between donor and acceptor sites is around 2300 cm$^{-1}$ 
and that the narrow spectral region about $E_{da}$  is giving rise to the observed ability
for Drosophela to discriminate between deuterated isotopomers by scent. 
It is possible that the  tunneling gap may be tuned to the nitrile (-C$\equiv$N) stretching mode which does lie in this spectral region
 and may serve a role in detecting a variety of naturally occurring nitrile containing odorants. 
A test of our model would be to identify an olfactant that has IR active C-H stretching modes but upon ionization distorts along modes that do not involve the C-H stretching modes. For example, 
it is known that upon ionization of ethylene, the C=C bond-length in $C_{2}H_{4}^{+}$ is 
shorter than in $C_{2}H_{4}$ and the C-H bond-lengths are unchanged.  
This suggests that the energy gradient in our model will be directed primarily 
along the C=C bond and have very little projection onto the C-H stretching modes.  
However, the IR spectrum does show an isotope shift of the CH stretching  
modes into the 2300cm$^{-1}$ region.  This is evident in our computed spectra shown in Fig.~\ref{ethylene}.  
Our model then predicts that {\em Drosophila} should {\em not} be able to easily discriminate between ethylene and 
any of its isotopomers.  The model also suggests that the isotope effect may not generalize between different classes of 
organic olfactants.  For example, here we focused upon aromatic olfactants which are good hole acceptors 
since they have electron-rich $\pi$-systems.  However,  the electron transfer dynamics through aliphatic system
would not likely involve  transfer of a positive charge and may well involve different olfactory receptors all together.


\begin{acknowledgments}
ERB was supported by the
National Science Foundation (CHE-1011894) and  Robert A. Welch Foundation (E-1334). 
The work by GR was supported by DARPA (N66001-11-1-4119). 
ERB jointly conceived the study with GR, developed the theoretical model and prepared the
manuscript.  AC performed the quantum chemical calculations.  GR and AM designed and performed the experiments.  
The authors have no competing financial interests.

\end{acknowledgments}

%


\end{document}